\documentclass[conference]{IEEEtran}
\IEEEoverridecommandlockouts
\usepackage{cite}
\usepackage{amsmath,amssymb,amsfonts}
\usepackage{algorithmic}
\usepackage{graphicx}
\usepackage{textcomp}
\usepackage{xcolor}
\def\BibTeX{{\rm B\kern-.05em{\sc i\kern-.025em b}\kern-.08em
    T\kern-.1667em\lower.7ex\hbox{E}\kern-.125emX}}
\begin{document}

\title{Revelation of Task Difficulty in AI-aided Education}

\author{\IEEEauthorblockN{1\textsuperscript{st} Yitzhak Spielberg}
\IEEEauthorblockA{\textit{Computer Science Department} \\
\textit{Ariel University}\\
Ariel, Israel}
\and
\IEEEauthorblockN{2\textsuperscript{nd} Amos Azaria}
\IEEEauthorblockA{\textit{Computer Science Department} \\
\textit{Ariel University}\\
Ariel, Israel}
}

\maketitle

\begin{abstract}
When a student is asked to perform a given task, her subjective estimate of the difficulty of that task has a strong influence on her  performance.
There exists a rich literature on the impact of perceived task difficulty on performance and motivation.
Yet, there is another topic that is closely related to the subject of the influence of perceived task difficulty that did not receive any attention in previous research - the influence of revealing the true difficulty of a task to the student.
This paper investigates the impact of revealing the task difficulty on the student's performance, motivation, self-efficacy and subjective task value via an experiment in which workers are asked to solve matchstick riddles.
 Furthermore, we discuss how the experiment results might be relevant for AI-aided education. 
 Specifically, we elaborate on the question of how a student's learning experience might be improved by supporting her with two types of AI systems: an AI system that predicts task difficulty and an AI system that determines when task difficulty should be revealed and when not.
\end{abstract}

\section{Introduction}
\par  
When a student (or worker) receives a certain task, her subjective estimate of the difficulty of that task has a strong influence on her performance.
Research has shown that perceptions of task difficulty are strongly correlated to both performance metrics such as the success rate and the solution time and psychological factors that influence performance such as motivation, interest, self-efficacy and subjective task value (attainment value, intrinsic value, utility value). 

\par 
Most commonly, the student's perception of the difficulty of a given task is obtained \emph{implicitly} from 
the description of the task, the background or setting in which the task was provided, the duration of time allotted for the task, or other information related to the task.
Yet, there also exist tasks, for which the objective (true) difficulty of the task may be available for a teacher. In this case, the teacher will face an important question: should she reveal the objective difficulty of the task to the student? Will the revelation of the objective task difficulty benefit the student or will it be harmful to her performance? 
\par 
The influence of perceived task difficulty on the student's performance has been studied extensively  \cite{ref1}. 
Yet, to the best of our knowledge, the costs and benefits of revealing the objective (true) task difficulty to the student have not been analyzed.
\par
\begin{figure}
	\centering
	\includegraphics[width=3.6in]{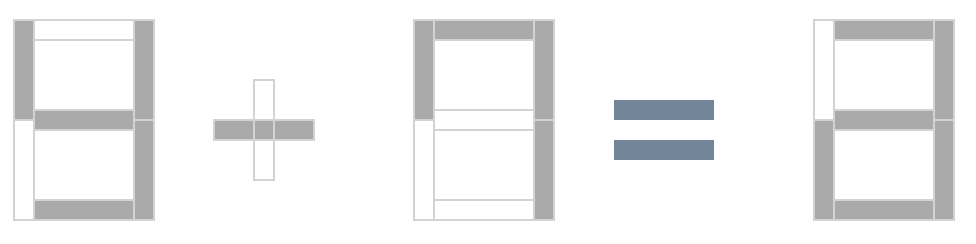}\\
	\caption {Example of a matchstick riddle where 2 matches need to be moved, in order to result in a legal mathematical equation.} 
	\label{rid_hard}
\end{figure}

 
This paper investigates the costs and benefits of revealing the true difficulty of a given task to the student that is supposed to solve that task. In particular, it examines the influence of revealing the task difficulty on performance metrics as well as psychological factors that impact performance, such as motivation, self-efficacy and subjective task value. We study these effects using the matchstick riddle task - a simple mathematical equation in which the numbers and the operator consist of matchsticks, some of which need to be moved in order to obtain a correct mathematical expression (see Figure \ref{rid_hard} for an example). 

\par 
For many tasks, the task difficulty is unknown a priori. However, for a given task category (such as matchstick riddles), it might be possible to build a data-driven AI that predicts the difficulty level of a given task from that category. 
Whether the development of such an AI is beneficial depends on the influence of revealing task difficulty on the student's performance. Therefore, understanding the influence of revealing task difficulty (in case that it is available) is particularly relevant with respect to AI-supported education.

\par 
Whereas the influence of revealing the task difficulty on the student's performance might be consistent (always positive or always negative), it might also be more complex. In particular, it might turn out that this influence depends on several external parameters, such as certain specific attributes of the task or the student's personality. For example, a student who enjoys challenges might become more motivated when she is told that the task is hard while revealing the same information to an anxious student might have the opposite effect. Therefore, in light of these considerations, it might be useful to consider an additional type of AI - an AI that predicts when to reveal the task difficulty and when not.

\par 
In summary, this paper investigates the following questions:
\begin{enumerate}
\item What is the influence of revealing the task difficulty on the student's performance, motivation, self-efficacy and subjective task value?
\item How to improve the learning experience using 2 types of AI systems - an AI system that predicts the task difficulty and an AI system that decides when to reveal the task difficulty?
\end{enumerate}

\section{Related Work}
\par  
As noted previously, to the best of our knowledge, there is no literature on the influence of revealing the difficulty of a given task on the student's performance. However, the topic of the impact of revealing the task difficulty is closely related to the topic of the impact of perceptions of task difficulty - a subject that has been discussed in a number of scientific publications. This section lists the central conclusions with respect to the influence of perceived task difficulty on performance, on the main psychological factors that influence performance (motivation and self-efficacy), and on subjective task value. One remarkable conclusion from the literature that has been reviewed is the existence of contradicting hypotheses - particularly, concerning the impact of perceived task difficulty on performance and motivation. Since the various studies have taken place on different tasks, one possible explanation for the contradictions might be that the relationship between perceived task difficulty and performance/motivation depends on characteristics that define the specific nature of the given task. The following discussion of the relevant literature is structured in accordance with the following four factors: performance, motivation, self-efficacy and subjective task value. 

\par 
\textbf{Performance}: In this paper, the performance of a cohort of students on a task is being measured by two metrics: the success rate and the average solution time. There exists a rich literature on the influence of perceived task difficulty on performance \cite{ref1}.
There are two main hypotheses that appear in the literature.
The first hypothesis states that the student's performance decreases as perceived task difficulty increases. According to the second hypothesis, however, the perceived task difficulty has no influence on the student's performance. Although there seems to be more support for the first hypothesis (\cite{ref1,ref2,ref5}), there is also some support for the second hypothesis \cite{ref4}. 
Although, on the first glance, these two hypotheses seem to contradict each other, it might be the case that upon deeper reflection there is no contradiction between them, because the studies were performed on different tasks and the relationship between perceived task difficulty and performance/motivation might depend on the specifics of the task.

\par 
\textbf{Motivation}: 
The literature on the influence of perceived task difficulty on motivation also contains contradicting hypotheses. The first hypothesis, namely, that motivation is negatively correlated to perceived task difficulty
is supported by Hom \cite{ref4}. Barron et. al. also provide evidence for this hypothesis and augment it with an additional claim; according to their analysis motivation is mediated by self-efficacy. They suggest that an increase in perceived task difficulty leads to a decrease in self-efficacy which causes a decrease in motivation \cite{ref9}. The contradicting hypothesis, that higher perceived task difficulty leads to higher motivation - appears in the literature as well. Boggiano et. al. suggest that not only the level of motivation but also the level of interest increases when students perceive a task as being more difficult  \cite{ref8}. If high levels of exertion are considered to be an expression of high levels of motivation then the finding that higher perceived task difficulty leads to higher levels of exertion \cite{ref11} can also be regarded as a support for the contradicting hypothesis. Another alternative hypothesis that can be found in the literature is that there is no relation between perceived task difficulty and motivation. Evidence for this alternative claim can be found in Robinson et. al. \cite{ref1a}.

\par 
\textbf{Self-Efficacy}: The term ``self-efficacy'' denotes the student's confidence in her ability to succeed in a given task. There exists a rich literature on the connection between perceived task difficulty, self-efficacy, motivation and performance (see for example \cite{ref5,ref5b,ref13}). Particularly, it is worth noting, that self-efficacy is closely related to motivation, such that higher self-efficacy usually leads to higher motivation. Therefore, it makes sense to assume that the relationship between perceived task difficulty and self-efficacy is very similar to the relationship between perceived task difficulty and motivation. Since the first hypothesis from the paragraph above - that higher perceived task difficulty leads to lower motivation - has the strongest support in the literature, it might be reasonable to suspect that there also exists literature that confirms the analogous statement with respect to self-efficacy - and, indeed, this seems to be the case (see \cite{ref3,ref3b,ref3c}). Concerning the relationship between high perceived task difficulty and low self-efficacy, it is particularly interesting that it is not clear which factor is the cause and which - the effect. While Wigfield and Eccles suggest that higher perceived task difficulty leads to lower self-efficacy \cite{ref3}, Li et. al. \cite{ref5} claim the reverse cause-effect relationship: a student is inclined to consider the task as more difficult precisely because she has low confidence in her ability to solve the task.

\par 
\textbf{Subjective Task Value}: The term ``subjective task value'' relates to the various ways in which a given task can be valuable to the student. Expectancy-value theory of motivation \cite{ref3} defines three types of subjective task values: 
\begin{itemize}
    \item Intrinsic value: the level of pleasure that the student derives from the task.
    \item Identity value: the amount of self-confidence gained from succeeding in the task.
    \item Utility value: the value of the knowledge and skills obtained from performing the task.
\end{itemize}
A review of the literature revealed that perceived task difficulty influences  all categories of subjective task value. With regard to intrinsic value, Li et. al. suggest that a student who perceives a task as very difficult will experience lower levels of pleasure in comparison to a student who perceives the same task as less difficult \cite{ref5}. With regard to identity value, the literature contains multiple hypotheses. While Brown \cite{ref10} presents evidence for a positive correlation between perceived task difficulty and identity value, Li et. al. \cite{ref5} do not find any correlation at all. With regard to utility value, we found only one hypothesis in the literature: Li et. al. suggest that utility value increases as perceived task difficulty increases \cite{ref5}.

\section {Experimental Setting}
\subsection{The Task}
 \par
\begin{figure}
	\centering
	\includegraphics[width=3.6in]{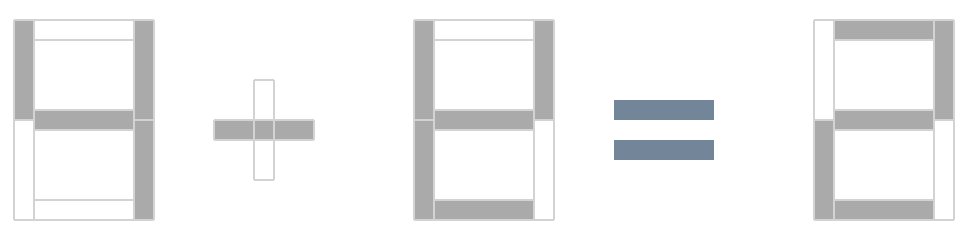}\\
	\caption {Example of a matchstick riddle where 1 match needs to be moved.} 
	\label{rid_easy}
\end{figure}

 
 In order to determine the effects of revealing the difficulty of a task to a student we use the matchstick riddle as our test-bed.
 A matchstick riddle is an equation with 4 tokens (3 digits, 1 operator) and an ``='' sign, in which each token consists of multiple matchsticks. The digit tokens can represent any digit 0,1..,9 and the operator token can be a `+' or `-'.  The riddle starts with an incorrect mathematical expression, such as `3+2=7' or an expression that contains invalid tokens. In order to solve the riddle, the participant needs to move 1 or more matchsticks to transform the expression into a correct mathematical equation. The amount of matches to be moved is displayed to the participant. Clearly, matchstick riddles have different difficulty levels, see Figure \ref{rid_hard} for an example of a difficult riddle, and Figure \ref{rid_easy} for an example of an easy riddle. The solution to Figure \ref{rid_hard} is ``$9-1=8$'', and the solution to Figure \ref{rid_easy} is ``$4-2=2$''.

 \par 
 We use the domain of matchstick riddles because the computer can automatically generate many tasks of different difficulty and easily check if the answer is correct. Furthermore, since this type of task is not well known (unlike mathematical questions, English questions or sudoku), we could expect the performance to be somewhat consistent between different participants, so using the performance of some participants to predict the difficulty of the task may seem more appropriate when using the matchstick riddle.
 
\par 
All experiments (described below) were performed on a collection of matchstick riddles that was generated by a program that we implemented. The collection consists of 500 riddles with 1 match to be moved and another 500 riddles with 2 matches to be moved. Although, as one might intuit, 2-stick riddles are more difficult than 1-stick riddles on average, it turned out that there were many 1-stick riddles that were more difficult than 2-stick riddles, according to the performance statistics. This observation suggests that the difficulty of a riddle depends not only on the number of matches that need to be moved but also on the specific locations of these matches.

\subsection{Experiment Details}
\par 
To investigate the influence of revealing task difficulty we conducted an experiment in which the participants (or workers) were asked to solve matchstick riddles. The experiment, which was implemented as a HIT (human intelligence task) on the Amazon Mechanical Turk platform, can be described by the following characteristics.
\begin{itemize}
    \item The worker saw the riddle and was informed about the number of matches that needed to be moved. To solve the riddle she needed to select the correct matches and to move them into the appropriate positions.
    \item The order in which the riddles were being displayed to the worker was random (without repetition).
    \item In order to provide some motivation for the worker, the worker received a small monetary reward for each correct solution.  
    \item The worker had an unlimited amount of time for each riddle. She could move on to the next riddle whenever she liked.
    \item The worker decided by herself how many riddles she wanted to complete. She was allowed to finish the HIT whenever she desired.
    \item The worker's performance was measured on each riddle. More precisely, the system measured two variables: the time spent on the riddle and whether the worker was able to solve the riddle or not. 
\end{itemize}
\par 
After completing the HIT the worker was asked to fill out a short survey. The survey contained the following statements:
\begin{enumerate}
\item I enjoyed the task. (S1)
\item I found the task interesting. (S2)
\item I believe that I performed well on the task. (S3)
\item I would like to perform similar tasks in the future. (S4)
\end{enumerate}

All statements had to be evaluated on a 7-level Likert scale, by choosing an answer from the following list: ``strongly disagree" (1), ``disagree" (2), ``somewhat disagree'' (3),``neutral" (4),``somewhat agree'' (5), ``agree" (6), and ``strongly agree" (7). 

\par 
The experiment consisted of two versions of the riddles HIT. In the first version, the riddle difficulty level was not displayed. In the second version, the difficulty level of the riddle was displayed on top of the riddle. In the following elaboration, the group of workers that performed the 1st/2nd version will be called group 1 and group 2 respectively. The group sizes were N1=97 and N2=125.

\par 
For each of the two groups the individual performance statistics were aggregated  into group-level performance statistics by averaging. The following group-level statistics have been computed:
\begin{itemize}
    \item Average attempt ratio, which equals the number of riddles attempted divided by the total number of riddles seen. A riddle is considered to be  'attempted' if the worker spent at least 20s on it.
    \item Average solution ratio, which is the number of riddles solved divided by the number of riddles seen.
     \item Average time spent on the HIT.
    \item Average solution time, which includes only solved riddles.
    \item Average riddle time, which includes all riddles (solved and unsolved).
\end{itemize}
All of these group-level statistics were computed not only on the collection of all riddles (table \ref{stats_all}), but also for the following two subsets of riddles: easy riddles (table \ref{stats_easy}) and hard riddles (table \ref{stats_hard}), as described hereunder. 
Furthermore, group-level statistics were computed for the survey results.

\subsection{Determining the Task Difficulty} 
\par 
As mentioned previously, for group 2 the difficulty level of the riddle was displayed on top of each riddle. The possible difficulty levels were 1,2,..,5 where 1 corresponded to 'very easy' and 5 to `very hard'. In the discussion of the results the scale that will be used will be a coarser scale with only 3 levels - for greater clarity. The category `hard' will correspond to levels 4 and 5 on the finer scale, the category `medium' will correspond to level 3 and the category `easy' will correspond to levels 1 and 2. 
\par 
The difficulty level of each riddle was specified before the experiment by running the HIT on a separate group of workers that did not participate in the experiment afterward.
The difficulty level of a riddle was determined by the average solution time of that riddle: the higher the average solution time the higher the difficulty level. 
The size of the group was determined such that each riddle was solved by approximately 10 workers. The function for mapping the average solution time to difficulty level assumes that the number of riddles in each difficulty category is roughly the same. That is, the average solution times are partitioned into five quantiles---one for each difficulty level, where the first quantile contains the lowest solution times (the easiest tasks, i.e., difficulty level 1) and the fifth quantile contains the highest solution times (difficulty level 5). 
\section {Experimental Results}
\par 
The purpose of the experiment was to understand the influence of revealing task difficulty on multiple performance-related features. This section compares the results of the two groups of workers (group 1 and group 2) with respect to performance and three performance-related features - motivation, self-efficacy and subjective task value. All statistics were computed on the entire collection of riddles and on two subsets: easy riddles and hard riddles. This means that each statistic comes in three versions. For example, there exists an average solution time on the set of all riddles, an average solution time on the set of easy riddles and an average solution time on the set of hard riddles. 

\begin{table}
\centering
\begin{tabular}{ l|c c  }
   & Group 2 & Group 1 \\
   & (difficulty displayed) & (w/o difficulty)\\
 \hline
 No. workers &  129  & 100  \\
 No. riddles seen & 4459 & 3443 \\
 Avg seen &34.57& 34.43 \\
 Avg solved &30.9& 30.28 \\
 Avg attempted &31.95&31.89 \\
 Avg solution ratio &0.89&0.88\\
 Avg attempt ratio &0.92&0.93\\
 Avg time on HIT &3238&2745 \\
 Avg solution time &91.92&77.61\\
 Avg riddle time &93.68&79.74\\
\end{tabular}
\caption{Performance statistics on the entire riddle collection.}
\label{stats_all}
\end{table}

\begin{table}
\centering
\begin{tabular}{ l|c c  }
   & Group 2 & Group 1 \\
   & (difficulty displayed) & (w/o difficulty)\\
 \hline
 No. workers & 114    & 89  \\
 No. riddles seen & 1796  & 1418  \\
 Avg seen & 15.75 & 15.93  \\
 Avg solved & 14.97 & 14.75  \\
 Avg attempted & 15.19 & 15.13 \\
 Avg solution ratio & 0.95 & 0.93 \\
 Avg attempted ratio & 0.96 & 0.95\\
 Avg time on HIT & 1014 & 892\\
 Avg solution time & 61.18 & 53.92\\
 Avg riddle time & 64.39 & 56.21\\
\end{tabular}
\caption{Performance statistics on easy riddles.}
\label{stats_easy}
\end{table}

\begin{table}
\centering
\begin{tabular}{ l|c c  }
   & Group 2 & Group 1 \\
   & (difficulty displayed) & (w/o difficulty)\\
 \hline
 No. workers & 118   & 86  \\
 No. riddles seen & 1756 & 1322  \\
 Avg seen & 14.88 & 15.37  \\
 Avg solved & 12.54 & 12.69  \\
 Avg attempted & 13.11 & 13.67 \\
 Avg solution ratio & 0.84 & 0.83\\
 Avg attempted ratio & 0.88 & 0.89 \\
 Avg time on HIT & 1845 & 1623\\
 Avg solution time & 127.11& 106.84\\
 Avg riddle time & 124.05 & 105.6\\
 \end{tabular}
 \caption{Performance statistics on hard riddles.}
\label{stats_hard}
\end{table}

\begin{table}
\centering
\begin{tabular}{ l|c|c  }
   & Group 2 & Group 1 \\
   & (difficulty displayed) & (w/o difficulty)\\
 \hline
 No. workers & 125    & 97  \\
 Enjoyment & 6.34  & 6.38  \\
 Interesting & 6.42  & 6.39  \\
 Perform well & 5.86 & 5.81  \\
 Similar in future & 6.53 & 6.54 \\ 
\end{tabular}
\caption{Survey results}
\label{stats_survey}
\end{table}

\subsection{Influence on Performance}
\par  
One of the key goals of this paper is to understand the impact of revealing the task difficulty on the student's performance. The 2 variables that we use to measure the performance, are the solution ratio and the average solution time. 
We are particularly interested in two questions: (a) Whether revealing that an easy task is easy improves the performance. (b) Whether revealing that a hard task is hard diminishes the performance. For this purpose, the performance statistics are measured and analyzed both on the entire set of riddles and on the two subsets of riddles (i.e., easy riddles and hard riddles).  
\par 
The analysis of the experiment results shows that there is no statistically significant difference between the two groups with regard to the solution ratio. This is true both for the entire collection of riddles and for the two subsets. The solution for group 1 and group 2 ratios are similar. For example, on the subset of easy riddles, the average solution ratio is 0.93 for group 1 and 0.95 for group 2.

\par  
However, with regard to solution time, it seems that the influence of revealing task difficulty is negative. On the subset of easy riddles, there is no significant difference between the groups. On the subset of hard riddles, however, group 1 clearly outperforms group 2, with 107 seconds for group 1, and 126 seconds for group 2. 
This result contradicts the initial hypothesis that expecting a task to be easy improves the performance, at least on the matchstick riddle task.

\subsection{Influence on Motivation}
\par  
Another important question is the influence of revealing task difficulty on  motivation. We use the average attempt ratio and the average riddle time as proxies for the worker's level of motivation. This approach is based on the intuition, that a worker who is more motivated will give up less quickly, which leads to a higher average riddle time as well as to a higher attempt ratio. 

\par 
The experiment results reveal that on the subset of hard riddles the average riddle time is higher for group 2 than for group 1 (106s (group 1), 124s (group 2)). This result holds for hard riddles only. On the entire collection of riddles and on the subset of easy riddles there is no significant difference in the average riddle time between the 2 groups. This indicates that in certain cases, such as the specific matchstick riddles task, knowing that a task is hard might increase the student's motivation and persistence.

\par 
With regard to the attempt ratio, there is no statistically significant difference between the two groups. This result holds both for the entire collection of riddles and each of the two subsets of riddles (easy riddles, hard riddles).

\par 
With regard to the survey statistics, the one statement from the survey that might be particularly relevant with respect to motivation is S2 (``I found the task interesting''), as interest facilitates motivation. The experiment results seem to indicate that those workers who were informed about the task difficulty did not find the task more interesting than those workers that were not. There is no significant difference between the 2 groups with regard to the score on S2 with an average of 6.39 for group 1, and 6.42 for group 2.

\subsection{Influence on Self-Efficacy}
\par 
The term 'self-efficacy' denotes a student's confidence in his or her ability to succeed in a given task. As noted above, there might be a close relationship between self-efficacy and motivation, such that, in most cases, a student who is highly motivated on a given task has high confidence that she will be able to succeed in it. Therefore, it might be assumed that the influence of revealing task difficulty on self-efficacy is very similar to the influence of revealing task difficulty on motivation.
\par 
While it is possible to measure motivation via performance metrics, such as the average attempt ratio, it is not possible to measure self-efficacy this way. Usually, the level of self-efficacy is measured by asking the student about her confidence in succeeding on the task before she starts the task. In our case, however, such an approach would have been infeasible, since in one HIT an average worker attempts about 30 riddles and it might cause her too much discomfort to ask about her confidence level before each riddle. For this reason, we use a different approach. Instead of measuring self-efficacy, we measure the a-posteriori version of self-efficacy: the worker's confidence about her performance on the task after completion. For this purpose, the statement S3 (``Do you believe that you performed well on the task?'' ) was included in the survey that was filled out by every worker after completing the HIT.

\par 
The analysis of the survey results (in particular the S3 statement) shows that indeed, the relation between revealing the task difficulty and self-efficacy is very similar to the relation between revealing the task difficulty and motivation. There is no significant difference between the 2 groups in the score on S3, with 5.81 for group 1, and 5.86 for group 2. This result is very similar to the corresponding result with respect to motivation. As discussed previously, those metrics that are indicative of the level of motivation (with the exception of riddle time for hard riddles) were not influenced at all by the revelation of task difficulty.

\subsection{Influence on subjective Task Value}

\par 
As mentioned, there are three types of subjective task values: intrinsic value, identity value and utility value. Although the statements in the survey do not precisely correspond to these three types of subjective task value, three of the four statements can be closely linked to them. In particular, it seems meaningful to associate intrinsic value with S1 ("Did you enjoy the task?"), while identity value and utility value might be associated with S2, S4, albeit more loosely.

\par 
In order to determine whether revealing task difficulty influences any of the types of subjective task values, the average scores on the corresponding survey statements have been analyzed. With respect to S1, the difference in the scores of 6.38 for group 1 and 6.34 for group 2, is not large enough to be statistically significant. The same conclusion holds true on S2 and S4 as well, where the scores of the two groups are also close to each other. These results reject the intuitive hypothesis that students who are informed about the task difficulty might assign a higher identity value to the task (for example, students might find it rewarding to solve a task when they know that it is hard).

\section{AI that predicts the Task Difficulty}
\par 
In the previous section, we have analyzed the influence of revealing task difficulty on the student's performance and some performance-related features such as motivation and self-efficacy. 
The analysis showed that, indeed, there are certain performance-related features
that are positively influenced by revealing the task difficulty, such as average riddle time on hard riddles.
In order to reveal the task difficulty to the student, the system (software) that feeds the tasks to the student needs to know the difficulty of every task from the given category of tasks (e.g. matchstick riddles).
Although for certain task instances the difficulty level might be known a priori from performance statistics, such as solution time and solution ratio,
there may be many other task instances whose difficulty level is unknown.
This calls for a difficulty predictor (an AI model) that would be able to compute the difficulty level of any task from a given task category automatically.  

\par 
Although it might be unrealistic to design a generic difficulty predictor that would operate on all task categories, it might be possible to create one that is specific to a given category of tasks, such as matchstick riddles. For any given category of tasks, this difficulty predictor would operate on a set of features that are specific to that task category. The difficulty predictor can be a model from an arbitrary class of predictor models such as support vector machines or deep neural networks. In case that the chosen class of predictor models is a neural network, the architecture of the network should be chosen with regard to the size of the training set, taking into account the general rule that larger networks require more training data.

\par 
In order to train the difficulty predictor,  
a training set of tasks needs to be generated.
This can be accomplished in a similar way as was done for the matchstick riddles collection (see ``Experimental setting'' section).
The first step would be the generation of a collection of tasks from the given category. Afterward, a cohort of workers should be asked to solve these tasks and certain performance statistics such as the solution times and the solution ratios should be recorded. In the final step, these performance statistics might be leveraged to determine the difficulty level of each task in the training set.

\par 
For many task categories predicting the difficulty level of a task might be much simpler than solving it since the former only requires training a difficulty predictor on task-difficulty-tuples, while the latter calls for the development of a solution algorithm for the given task category, which might be extremely hard or even impossible for certain task categories. Therefore, it is important to emphasize that the task difficulty predictor does not need to know the solution of the task.
This is a very important feature of the predictor because it allows to build a predictor even when a solution algorithm is not available.

\section {Predicting Task Difficulty on the Matchstick Riddles Task}
\par 
 To test the ideas presented in the previous section we built a solution time predictor for the matchstick riddles task. Since the difficulty of a riddle, as we defined it, was determined only by the solution time, the solution time prediction can be transformed into a difficulty prediction. 
 Because of the staggering success of deep learning in various domains in recent years, the function class that was chosen for the predictor was the class of fully connected neural networks with rectified linear unit (ReLU) neurons.
In order to obtain optimal performance, multiple network architectures were tested - all of them consisting of 2-4 hidden layers and 10-20 neurons per layer.
 Furthermore, each layer was regularized by a dropout layer with a dropout rate of 0.5. The model training was performed using an early stopping strategy (the training session ended when the validation error did not decrease for a period of 10 consecutive epochs).

\par 
The basis for the data, that was used for training, validating and testing the model was a collection of 1300 matchstick riddles and the corresponding solution times. The experiment yielded 5-10 solution times for every riddle (each solution time corresponds to a different worker).
The median of these solution times was chosen as the prediction target. 
Thus, each sample in the resulting data set consisted of the riddle's feature vector and the corresponding median solution time.
We chose to predict the median solution time rather than the average solution time, because the median is more robust to outliers than the average. 
Finally, the data was partitioned into a training set (60 \%), a validation set (20 \%) and a testing set (20 \%).

\par 
In order to understand how the choice of features influences the quality of the difficulty predictor, multiple difficulty predictors have been trained - each one on a different set of features. The first predictor leveraged a minimal collection of features that used only the information contained in the riddle, but not the solution. This feature collection consisted of the match positions of the original expression (encoded as a one-hot vector) and the number of matches that needed to be moved. The second predictor was based on an extended feature collection that utilized the task solution. This extended collection consisted of the minimal feature collection, augmented by a one-hot vector, that encoded the initial and final positions of those matches that needed to be moved and a boolean feature that indicates whether there is a change in the operator ('+' to '-' or '-' to '+').
\par 
After training, all models were evaluated on the test set. 
With regard to the model performance, the differences between the different model architectures that were tested were negligible.
Moreover, it was observed that adding the solution-based features to the minimal feature collection did not improve the model performance.
All models, including those that used the extended feature collection, yielded a mean average error (MAE) of approximately 35 seconds on the test set. 
To evaluate the predictive power of the models, they were compared to a primitive model that produces a constant output for every riddle: the average target value (averaged over the training set). 
This primitive model produced a MAE of 42 seconds. Although this result is worse than the MAE of the neural network, unfortunately, the differences are quite small. 

\par 
There might be multiple explanations for the relatively poor performance of the neural network solution time predictor on the matchstick riddles task. 
One possible explanation for this phenomenon is the complexity of the given prediction task. It is not clear which features determine the complexity - aside from the number of matches that need to be moved.
Initially, we hypothesized that 2 features that are likely to make a riddle more difficult are a change in the operator and a change in the right side.
Yet, the analysis of the data does not support this hypothesis.
Furthermore, it might be worth noting that the prediction task (predicting the median solution time) cannot be accomplished by a human within a few seconds - a fact that might be another indicator of the complexity of the prediction task.

\section{AI that decides when to reveal the Task Difficulty}
\par 
The analysis of the experiment results showed, that revealing task difficulty indeed does have an impact on the student's performance and motivation, albeit only on certain subsets of tasks (for example on hard riddles).
Yet, from the experiment results, it did not become clear whether this influence is positive or negative. As reviewed above, the literature on the influence of perceived task difficulty contains contradicting hypotheses.
One possible interpretation of this finding might be the hypothesis that this influence might depend on additional factors that were not taken into account, such as certain specifics of the task or the student's attributes.
For instance, an anxious student might easily lose motivation by being told that the task is hard, while a very self-confident student, in contrast, might gain additional motivation from this information.

\par 
The assumption that the influence of revealing task difficulty depends on external factors, such as specific of the given task category or the student's personality, calls for an AI system that would determine in which cases task difficulty should be revealed.
 The inputs for this system might include various features, such as scores for the student's personality traits and possibly certain features of the given task category. The system's output could be either binary ('yes' or 'no') or continuous (a score that indicates the benefit of revealing the task difficulty).
 A more sophisticated system could produce scores that would reflect how beneficial revealing task difficulty would be with regard to every single performance-related feature: performance, motivation, self-efficacy and subjective task value. It should be mentioned that at this point the description of this system is rather abstract because more research would be required to understand how to design such a system. In particular, more research would be required to understand which factors modulate the influence of revealing task difficulty, how exactly they modulate this influence and how they correlated with each other. 

\par 
There exists evidence that confirms that a student's level of motivation on a given task depends on her personality \cite{ref12}. Therefore, the student's personality profile could be a key factor that determines whether revealing the task difficulty has a positive or negative impact on her motivation, which means that a model that determines in which cases task difficulty should be revealed, might require the student's psychological profile.
This psychological profile should contain scores on various personality traits, such as trait neuroticism and trait openness, which might be relevant with regard to the influence of perceived task difficulty. The field of psychometrics offers numerous ways to generate such a personality profile, among others, specialized psychological tests and questionnaires.

\section{Discussion \& Conclusion}

\par 
The current paper explored the influence of revealing the difficulty level of a task on the student's performance and several performance-related features such as motivation, self-efficacy and subjective task value.
The purpose of this effort was to understand whether revealing task difficulty improves the student's performance and to discuss how to support a student with two types of AI systems: an AI that predicts the task difficulty and an AI that determines when to reveal the task difficulty.
The matchstick riddles experiment produced several interesting results. While some of these results are intuitive and in line with the existing literature, others are less expected.

\par 
Three important experiment results, which are in line with the literature, can be regarded as intuitive. 
The first result is related to the solution time, which is an indicator of performance.
It has been observed that the solution time increased when the worker was told that the riddle is hard. This result is in line with the hypothesis from \cite{ref1,ref2,ref5} that states that an increase in perceived task difficulty causes a decrease in performance.
The two other notable results are linked to the influence of revealing task difficulty on motivation and interest.
With regard to these two features (motivation and interest), the experiment results indicate that revealing the difficulty of the riddles did not influence neither motivation nor interest. 

\par 
 Another result, related to performance, contradicts certain findings from the literature \cite{ref1,ref2,ref5}.
 Namely, when the worker was informed that a task is easy, her performance did not improve - neither with respect to the solution time nor with respect to the solution ratio.
 
 \par 
With regard to motivation, there are 2 counterintuitive experiment results that are also in contradiction to certain results from the literature  \cite{ref1a} .
 Firstly, it was observed that the worker's motivation did not decrease when she was informed that a riddle is hard---there was even evidence for an increase in motivation. Secondly, when a worker was informed that a riddle is easy, her motivation did not increase.

\par 
In conclusion, the riddles experiment showed that the impact of revealing task difficulty is neither clearly positive nor clearly negative, but depends on different factors such as the specific performance-related feature that is being inspected (performance, motivation etc.) and the difficulty of the task. 
We believe that there are four major conclusions that can be drawn from the experiment:
\begin{itemize}
\item \textbf{ Conclusion 1}: Revealing task difficulty has positive as well as negative effects on the student's performance and performance-related features (motivation etc.).
 \item \textbf{Conclusion 2}: In some cases, the impact of revealing task difficulty can be ambiguous (positive and negative simultaneously). For example, revealing task difficulty on hard tasks leads to an increase in solution time, which can be regarded either as a decrease in performance or as an increase in exertion.
 \item \textbf{Conclusion 3}: Revealing the task difficulty seems to impact only some of the performance-related features (performance and motivation) but has no impact on others (self-efficacy, subjective task value).
 \item \textbf{Conclusion 4}: The impact of revealing task difficulty might depend on the specifics of the task and the student's personality.
\end{itemize}

\par 
It is important to emphasize that the experiment discussed in this paper was performed on a single task category (matchstick riddles). Thus, it is not certain that the conclusions, derived from the experiment results can be generalized to other task categories. For this reason, an interesting direction of research for future work could be to perform a similar experiment on other task categories (e.g. sudoku).

\par 
 The goal of the paper was to investigate the utility of supporting students with an AI that predicts task difficulty for a given task category (such as matchstick riddles). The analysis of the experiment results showed that in many cases revealing task difficulty can have positive effects.
 Therefore, an AI that predicts task difficulty might be useful for increasing the students' performance and motivation.
 On the other hand, the experiment results also indicate that sometimes revealing task difficulty impacts the student in a negative way.
 For this reason, the difficulty predictor should be augmented with an AI that determines when to reveal task difficulty and when not to do so.

\section*{Acknowledgment}
This research was supported in part by the Ministry of Science, Technology \& Space, Israel. 

\bibliographystyle{plain}  
\bibliography{myrefs.bib}  

\end{document}